\documentclass[usenatbib]{mn2e}
\usepackage{epsfig,rotating,natbib,amssymb}

\title{Toy Stars in two dimensions}

\author[Monaghan \& Price]{J.J. Monaghan$^1$, D.J. Price$^2$ \\
$^1$School of Mathematical Sciences, Monash University, Clayton 3800, Australia\\
$^2$School of Physics, University of Exeter, Exeter EX4 4QL, UK \\
}
\pagerange{\pageref{firstpage}--\pageref{lastpage}} \pubyear{2005}
\date{submitted: 1st August 2005 Accepted: 25th October 2005}

\begin{document}
\label{firstpage}
\bibliographystyle{mn2e}
\maketitle

\begin{abstract}
 Toy Stars are gas masses  where the  compressibility is treated without approximations but gravity is replaced  by a force which, for any pair of masses, is along their line of centres and proportional to their separation.  They provide an  invaluable resource for testing the suitability of numerical codes for astrophysical gas dynamics.   In this paper we derive the equations for both small amplitude oscillations and non linear solutions for rotating and pulsating Toy Stars in two dimensions, and show that the solutions can be reduced to a small number of ordinary differential equations.  We compare the accurate solutions of these equations with SPH simulations. The two dimensional Toy Star solutions are found to provide an excellent benchmark for SPH algorithms, highlighting many of the strengths and also some weaknesses of the method.
\end{abstract}
   

\section{Introduction}
\setcounter{equation}{0}

A fundamental computational problem in astrophysics is the motion of a cloud of gas forming a protostar in an ambient medium which is typically lower in density  by a factor of $< 10^{12} $.  The ambient medium can therefore  be treated as a vacuum to an excellent approximation.  The  simulation of such a gas cloud is  more difficult than many of the standard battery of test problems considered in computational astrophysics.   Exact, or very accurate, solutions for the motion of these gas clouds are rare and this creates further difficulties in assessing the accuracy of a numerical algorithm.  

In this paper we fill this gap in test cases by studying  solutions and simulations for the class of models which we have called Toy Stars.  They are, first and foremost, a tool for studying the accuracy of computational algorithms relevant to astrophysical gas dynamics but they also provide an interesting class of dynamical systems where exact solutions can be found.  The fundamental aspects of Toy Stars and a variety of one dimensional solutions have been discussed by \citet{mp04}.  The key features are that they are gas masses where compressibility is included without approximation, but gravity is replaced by a force derived from the potential 
 \begin{equation}
 \Phi =  \frac{\nu}{4} \sum_{i=1}^N \sum_{j=1}^Nm_i m_j ( {\bf r}_i - {\bf r}_j )^2,
 \end{equation}
 where $\nu$ is a constant, $m_i$ is the mass of particle $i$ and it is assumed that there are $N$ masses.  The equation of motion of particle $j$ in the absence of other forces is
  \begin{equation}
 \frac{d {\bf v}_j}{dt} = - M \nu {\bf r}_j,
 \end{equation}
 where $M$ is the total mass and the origin is the centre of mass.  Remarkably, as first noted by Newton \citep[see][]{chandra95}, the particles move {\it  independently} about the centre of mass of the particle system.  Two particles in  a binary Toy Star system move on orbits given by the solutions of the following equation for the relative coordinate ${\bf r = r_1 - r_2}$
 \begin{equation}
\frac{d^2 {\bf r} }{dt^2}  =  - M \nu {\bf r} ,
\end{equation}
These solutions are closed Lissajous figures which include elliptical orbits. 
 
 The equations of motion of a gaseous system in a Toy Star force are the acceleration equation
  \begin{equation}
  \frac{d {\bf v}}{dt} =    - \frac{1}{\rho} \nabla P  - \Omega^2 {\bf r}
  \end{equation}
  where $\Omega^2 = M \nu $,  $P$ is the pressure and $\rho$ the density, and the continuity equation
  \begin{equation}
  \frac{d \rho}{ dt} = -\rho \nabla \cdot {\bf v}.
  \label{eq:cty}
  \end{equation} 
  If $P = K \rho^2$ this equation together with the continuity equation is identical to the equation for small oscillations of water in a lake with paraboloidal bottom.  This equation has been studied by \citet{gold30} and \citet{lamb32}, but the most important contributions were made by \citet{ball63,ball65} and \citet{holm91}.  In this paper we focus on the aspects of these equations which are most important for astrophysical problems, and generalise to the case where $P = K \rho^\gamma$.  
  
  In the following we first consider the small amplitude oscillations of the Toy Star and compare the results to simulations using SPH.  We then analyze the non linear motions for arbitrary $\gamma$ and show that the equations reduce to a small set of ordinary differential equations which can be integrated with high accuracy.  We then compare the results of these integrations with SPH simulations.

\section{Static structure and Oscillations}
If $P = K \rho^\gamma$ the density of the  static model is given by
\begin{equation}
\rho^{\gamma -1} = \rho_0^{\gamma -1} \left ( 1 - \frac{r^2}{r_e^2}  \right ),
\end{equation}
where the radius $r_e$ is 
\begin{equation}
r_e^2 = \frac{ 2 K \gamma \rho_0^{\gamma-1}  }{ \Omega^2 (\gamma-1) }.
\end{equation}
The mass $M$ is given by
\begin{equation}
M = 2 \pi \int_0^{r_e} \rho r dr = \frac{\pi r_e^2 (\gamma-1) \rho_0}{\gamma}.
\end{equation}

If the  $|{\bf v}| \le c_s$, where $c_s$ is the unperturbed speed of sound,  the equations of motion can be linearized about the static structure. It is convenient to define 
\begin{equation}
R = \rho^{\gamma-1}
\end{equation}
and write $R = \bar{R} + \eta$ where $\bar{ R}$ is $R$ calculated using the unperturbed density $\bar{\rho}$.   The equations of motion then become
\begin{eqnarray}
\frac{\partial {\bf v}}{\partial t}  & = & - \frac{K \gamma}{\gamma-1 }  \nabla \eta, \label{eq:dvdtgradeta} \\
\frac{\partial \eta}{\partial t} & = & - {\bf v} \cdot \nabla \bar{R} - (\gamma-1) \bar{R} \nabla \cdot {\bf v}. 
\end{eqnarray}
We assume the time variation is $e^{\imath \sigma t }$, and write
\begin{eqnarray}
\eta & = & D e^{\imath \sigma t} ,\\
v & = & V e^{\imath \sigma t}.
\end{eqnarray}
If these expressions are substituted into the linearized equations,  $V$ can be eliminated to get the following equation for $D$
\begin{equation}
\left ( 1 - \frac{r^2}{r_e^2}  \right ) \nabla^2 D - \frac{2r}{r_e^2 (\gamma-1)} \nabla D + \frac{\sigma^2 }{ K \gamma \rho_0^{\gamma-1} } D = 0.
\end{equation}
Assuming separable solutions, they must be of the form $\zeta(r) \sin{\theta} $ or $\zeta(r) \cos{\theta} $ and the equation for $\zeta$ is
\begin{eqnarray}
\left ( 1 - \frac{r^2}{r_e^2}  \right )  \left (  \frac{d^2 \zeta}{dr^2} + \frac{1}{r}  \frac{d \zeta}{dr} - \frac{s^2 \zeta}{r^2}  \right ) & & \nonumber \\
  - \frac{2r}{r_e^2 (\gamma-1)} \frac{d \zeta}{dr}+ \frac{\sigma^2}{K \gamma \rho_0^{\gamma-1} } \zeta & = & 0.
\end{eqnarray}
The solutions of this equation determine the values of $\sigma$.   While this equation can be transformed to the equation for a Hypergeometric function it is more convenient to determine the solutions directly using expansions in series following the method of Frobenius.  We thus take
\begin{equation}
\zeta(r) = X^c \sum_{n=0}^\infty a_n X^n,
\end{equation}
where $X=r/r_e$.  It is convenient to replace $\sigma$ by $\nu$ according to 
\begin{equation}
\nu^2  = \frac{ \sigma^2 r_e^2}{K \gamma \rho_0^{\gamma-1} } = \frac{2 \sigma^2}{\Omega^2 (\gamma-1) },
\end{equation}
using the definition of $r_e$. If the series is substituted into the equation for $\zeta$ we get the following recurrence relation for the coefficients

 \begin{equation}
a_{k+2 } = a_k \frac{ (k^2 + 2ks + 2(k+s)/(\gamma-1) - \nu^2)}{ (k+2+s)^2 - s^2 }
\label{eq:recurrencerelation}
\end{equation}
 
The indicial equation gives $c=s$ and there is one solution with $a_0$ arbitrary and $a_1$ zero.  Because the equation is second order there must be two arbitrary constants but because the solutions of the indicial equation differ by an integer (s is an integer) or are equal (s=0), the second arbitrary constant multiplies a solution containing $\ln{x}$, and must be zero.  The remaining series only converges if 
\begin{equation}
\frac{2 \sigma_j^2}{\Omega^2 (\sigma-1)} = (j+s ) \left (j+s + \frac{2}{\gamma-1} \right ) -s^2,
\end{equation}
where $j$ is an integer and the associated value of $\sigma$ is denoted by $\sigma_{j}$.
The last term in the series for a given $j$ is $a_j X^j$.  For numerical work we write the velocity in the form
\begin{eqnarray}
{\bf v }& = & {\bf V }\cos{(\sigma t)} \\
\eta & = & D \sin{(\sigma t)}.
\end{eqnarray}
From (\ref{eq:dvdtgradeta})
\begin{equation}
{\bf V} = \frac{K \gamma}{\sigma(\gamma-1)} \nabla D. \label{eq:vperturb}
\end{equation}
Further details of the linear modes are given in Appendix~\ref{sec:linearmodeappendix}.


\section{SPH simulations of static structure and linear oscillations}
 Smoothed Particle Hydrodynamics (SPH)\citep[for a recent review see][]{monaghan05} is a Lagrangian particle method for solving the equations of fluid dynamics. Since a primary application of SPH is to self-gravitating gas in astrophysical systems, in most cases involving free boundaries, Toy Stars represent an ideal test of the algorithm's capabilities on these systems. Whilst the standard SPH algorithm \citep[as described, for example in][]{monaghan92} has been well tested and benchmarked, we use the opportunity provided by the Toy Star solutions to benchmark more recent improvements to the algorithm. 

In particular we formulate the SPH equations from a variational principle such that the spatial variation of the smoothing length according to the density variation is accounted for self-consistently \citep{sh02,monaghan02,pm04b,price04}. We also use the Toy Stars to test a reversible time integration algorithm for SPH described in \citet{monaghan05}.  The specific implementation of the SPH algorithm used for the test problems presented in this paper is described below.

\subsection{SPH implementation}
\label{sec:sph}
 The SPH equations are formulated from a variational principle which accounts for the spatial variation of the smoothing length with density. Prior to the force evaluation, the density is calculated via a direct summation over the particles which is iterated to self-consistently determine both the smoothing length and the density according to the relation
\begin{equation}
h = \eta \left( \frac{m}{\rho} \right)^{1/2},
\label{eq:hrho}
\end{equation}
where $\eta$ is a constant relating the smoothing length to the average particle spacing. The procedure for doing this is described in \citet{pm04b} and \citet{price04}. Enforcing this relation tightly has been found to significantly sharpen the resolution of a typical SPH simulation and in conjunction with the additional terms in the momentum (and energy) equation(s), in general leads to substantial improvements in accuracy (although perhaps, as we will demonstrate in this paper, with a loss of robustness). 
 
 The pressure is calculated directly from the density using a polytropic equation of state $P = K \rho^{\gamma}$ and the standard cubic spline kernel \citep{monaghan92} is used.
 
  Damping towards the equilibrium solutions is achieved by applying a form of the SPH artificial viscosity used for shock capturing \citep{monaghan97} together with a damping in the force equation which is independent of resolution, given by
\begin{equation}
\frac{d{\bf v}}{dt} = - 0.02{\bf v} + {\bf f},
\end{equation}  
where ${\bf f}$ is the force per unit mass. Note that since the pressure is calculated directly from the density, the kinetic energy removed by the artificial viscosity and damping terms is not deposited as thermal energy but rather removed from the system. 

 In the linear and non-linear oscillation solutions to be described, a small amount of dissipation is applied selectively using the artificial viscosity switch described by \citet{mm97}. This switch turns on the viscosity terms in response to the magnitude of any negative divergence (ie. convergence) in the velocity field. In the oscillation solutions this artificial viscosity is applied only to approaching particles.

  The time integration is achieved using a manifestly reversible, second order integrator described by \citet{monaghan05} which in particular is reversible in the case of a variable step size (as for example when the step size is determined from the Courant condition). At the present time, the reversibility condition only holds with the density calculated by direct summation and the pressure determined directly from the density, which is sufficient for the calculations presented here (although it is fairly straightforward to generalise the integrator to the case in which the continuity and energy equations are also evolved). The differences between using the reversible integrator and a simple predictor corrector method as commonly used in SPH are found to be minor, although in the course of the Toy Star tests we have found several aspects of the reversible integrator which must be treated with caution.

\subsection{Static structure}
\label{sec:static}
 The simplest test case for the two dimensional Toy Star is to verify the static structure. This can be done in one of two ways: using equal mass particles which are damped into an equilibrium configuration or by varying the particle masses according to the equilibrium density profile. In order to test the robustness of our algorithm we examine both methods. The calculations are performed using an equation of state $P = \frac14 \rho ^{2}$.
\begin{figure}
\begin{center}
\begin{turn}{270}\epsfig{file=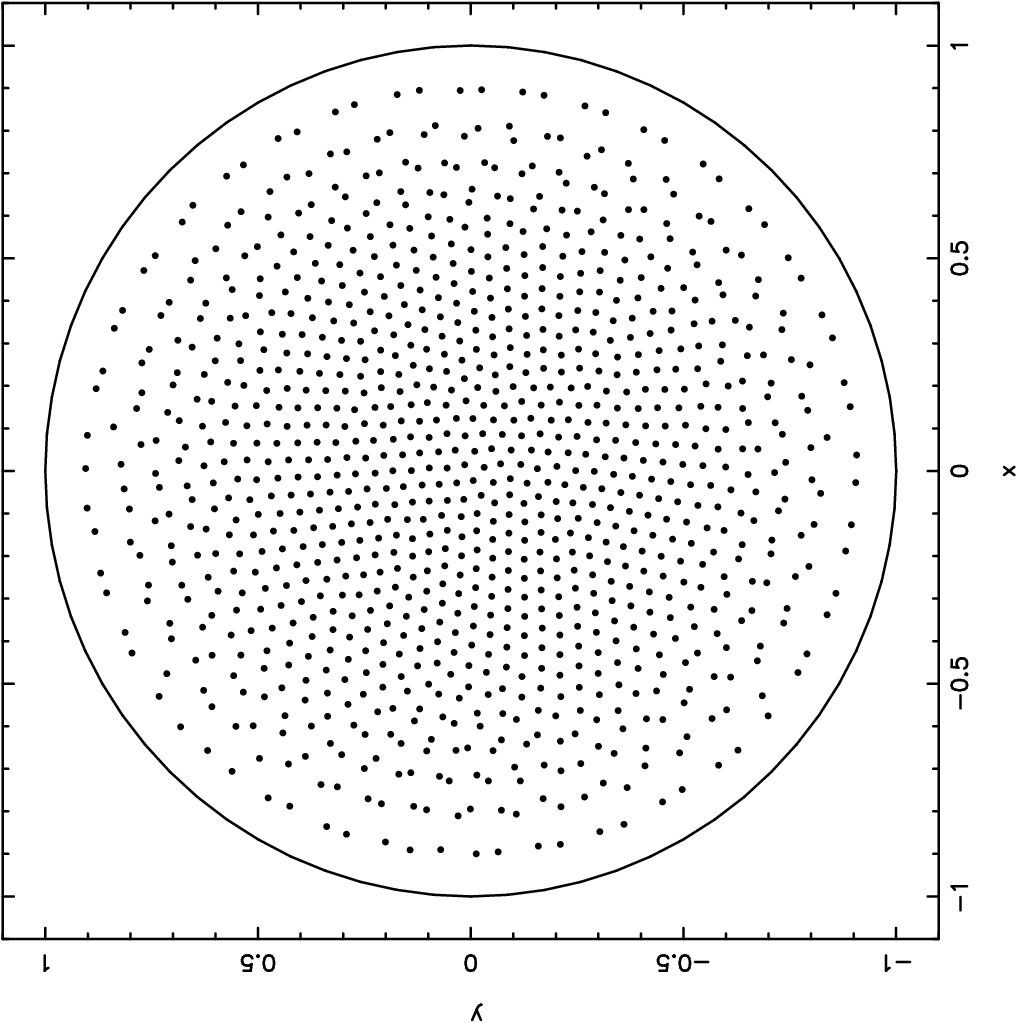, height=\columnwidth}\end{turn}
\caption{Toy Star static structure. We place 1000 equal mass SPH particles on a lattice of square cells within the circle of unit radius and allow them to evolve under the influence of the linear force.
The SPH particles are shown by the solid points after damping to an equilibrium
distribution whilst the solid circle shows the radius of the exact solution. The particles adopt a hexagonal lattice arrangement in the central regions and a ring-like arrangement at the outer edges}
\label{fig:static_eqm}

\begin{turn}{270}\epsfig{file=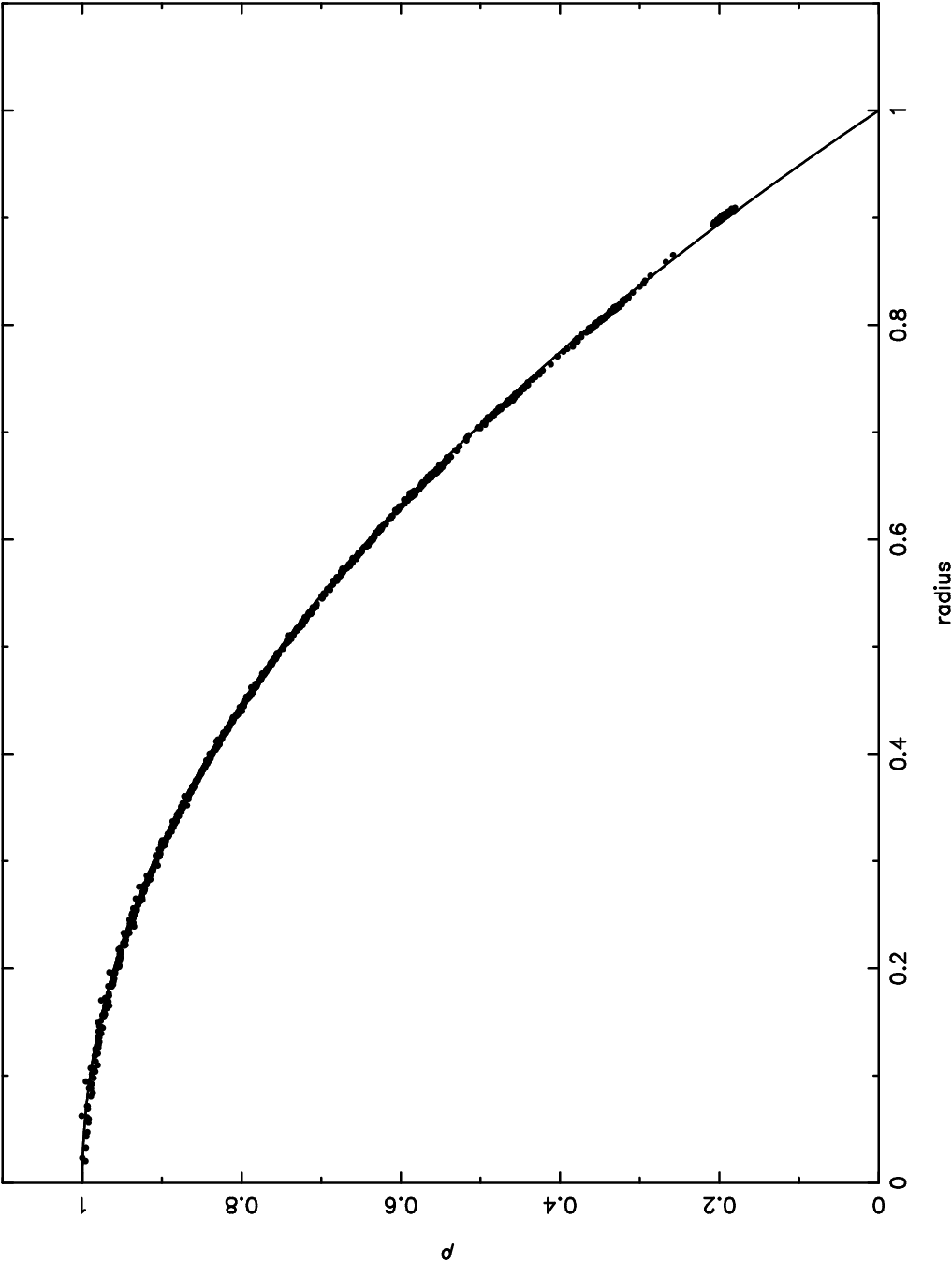, height=\columnwidth}\end{turn}
\caption{Density profile in the Toy Star static structure with equal mass particles. 
The SPH particles are shown by the solid points after damping to an equilibrium
distribution. The exact quadratic ($\rho = 1-r^2$) solution is given by the solid line.}
\label{fig:static_eqm_rho}
\end{center}
\end{figure}

\subsubsection{Equal mass particles}
  In the equal mass particle case, we construct the initial conditions by placing the particles on a lattice of square cells which is cropped to retain only those particles with a radius less than unity (chosen since it is the equilibrium radius of the Toy Star solution). Using a lattice spacing of $0.05$, this results in 1256 particles. The particles are perturbed from the lattice with a random amplitude of up to $\frac12$ of the initial lattice spacing to remove any residual effects from the initial regular particle arrangement in the equilibrium solution.

  The system is then allowed to collapse under the influence of the (axisymmetric) Toy Star force and damped until an equilibrium is reached (where typically we damp until $E_{kin} \sim 10^{-10} E_{tot}$). The equilibrium particle configuration is shown in Figure~\ref{fig:static_eqm} (where for comparison the solid circle on this plot marks the exact solution radius). The particles can be seen to adopt a reasonably isotropic close-packed hexagonal lattice arrangement in the central regions of the Toy Star, whilst at the edge tend to form discrete rings.  Experimenting with different initial particle configurations (including a completely random configuration) and different particle numbers result in the same distinguishing features (although with a particular caveat which is discussed below). The configuration adopted by the particles is in general related to the shape of the interpolation kernel used in the calculations. Thus the isotropic nature of the particle arrangement in the central regions results from the isotropy of the cubic spline kernel. The density profile of the equilibrium configuration is shown in Figure~\ref{fig:static_eqm_rho} and agrees well with the exact solution given by the solid line. The discrete rings of particles formed at the outer edges are also visible.

   The equal particle mass also provides a useful illustration of the `particle pairing' instability which occurs for larger smoothing lengths, specifically where the constant of proportionality in (\ref{eq:hrho}), $\eta > 1.2$ (where $\eta > 1.5$ is found to be particularly unstable). In these cases the particles tend to initially form pairs, which then begin to coalesce. Eventually these pairs clump together at the same location, with the resulting lattice adopting a close packed configuration similar to that shown in Figure~\ref{fig:static_eqm}  (albeit at half the resolution since two particles effectively become one). This is a well-known defect of the cubic spline kernel, resulting from the fact that the force tends to zero at the origin \citep[see for example][]{tc92,sha95}. Using $\eta > 1.5$ results in the closest neighbour being placed on the turning point of the force curve (in the isotropic, equilibrium configuration), thus moving inwards (and therefore clumping) as the particles are compressed.

 We have also investigated this instability using the standard SPH formalisms \citep{hk89,bea90} whereby a simple average of either the particle smoothing lengths or the kernel gradients is used in the equation of motion. In these cases particles are still observed to clump, although the lattice configuration tends to retain defects due to the averaging procedure. With the new variable smoothing length formalism the particles thus appear to settle more readily into minimum energy states because of the smoother lattice configuration (Figure~\ref{fig:static_eqm}). This means that the pairing instability is slightly more pronounced in this case (although we have not made a quantitative assessment).
 
  A solution to this instability may be easily obtained by using a kernel where the force does not decrease towards the origin. However, the density estimate using such kernels is in general quite poor.
 \citet{tc92} suggest a compromise approach in which the cubic spline kernel gradient is modified to remain constant in the region between the usual minimum and the origin whilst the unmodified cubic spline kernel is used in the density evaluation. Whilst this approach seems promising, it also introduces additional problems (for example, energy is not conserved exactly and the kernel gradient is no longer normalised). In this paper we simply use $\eta = 1.2$ in all of the simulations, thus avoiding the instability, although it is our intention to investigate the issue further elsewhere (in particular to examine alternative kernels to the cubic spline which have finite gradients at the origin but which do not lead to a significant decrease in interpolation accuracy).
\begin{figure}
\begin{center}
\begin{turn}{270}\epsfig{file=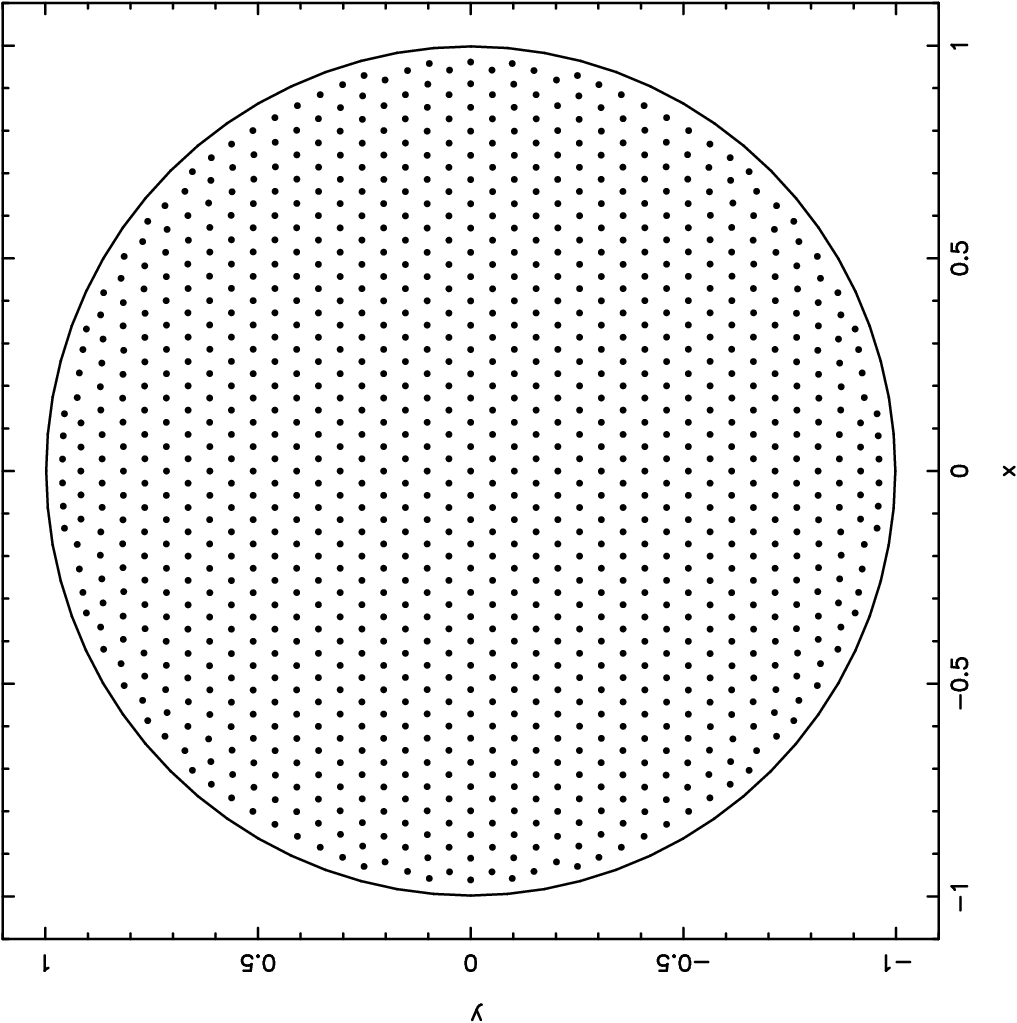, height=\columnwidth}\end{turn}
\caption{Toy Star static structure with unequal particle masses. 1045 SPH particles are set up in a hexagonal close packed lattice arrangement with their masses set according to the equilibrium Toy Star density profile. The particles are then allowed evolve under the influence of the linear force.
The SPH particles are shown by the solid points after damping to an equilibrium
distribution, whilst the solid circle shows the radius of the exact solution. }
\label{fig:static_varm}

\begin{turn}{270}\epsfig{file=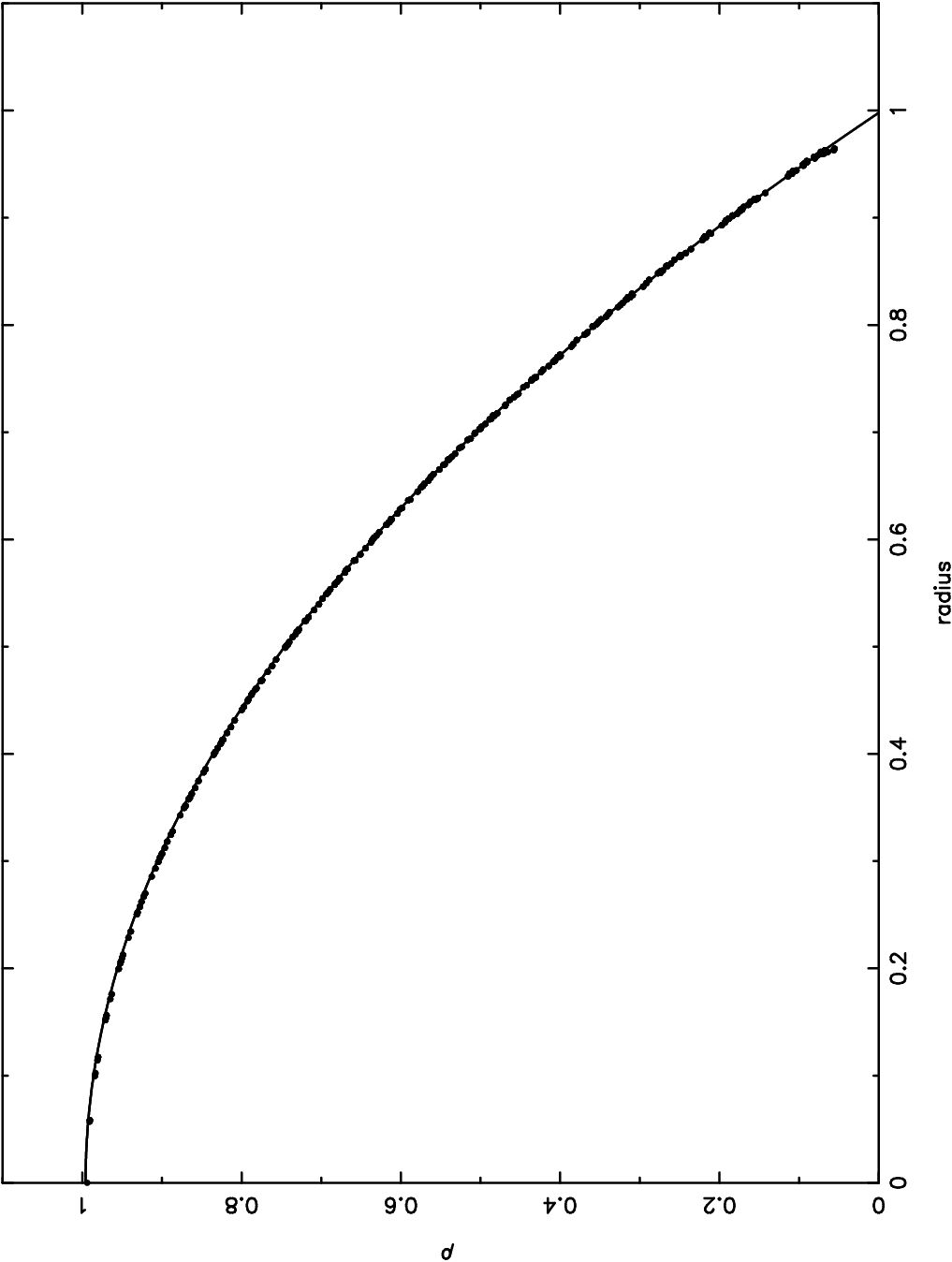, height=\columnwidth}\end{turn}
\caption{Density profile in the Toy Star static structure with 1045 unequal mass particles. 
The SPH particles are shown by the solid points after damping to an equilibrium
distribution. The exact quadratic ($\rho = 1-r^2$) solution is given by the solid line.}
\label{fig:static_varm_rho}
\end{center}
\end{figure}

\subsubsection{Unequal particle masses}
 In the second case, the particle masses were varied according to the equilibrium density profile. As a first attempt the particles were arranged on a lattice of square cells, with masses corresponding to the density profile at their radial position. To avoid problems with very low mass particles at the outer edge we excluded particles within half of the initial lattice spacing of the equilibrium radius ($r=1$ in this case).
 Allowing the particle distribution to evolve under the influence of the linear force, with damping applied as described above, the particles at the outer edges of the configuration at first shift slightly in order to give a more circular edge (rather than the stepped edge given by the cropped lattice).  When using the iterated smoothing length update the particles were then observed to gradually shift from the square lattice and attempt to adopt a hexagonal-type arrangement. An equilibrium configuration which is mutually satisfying for all of the particles then becomes very difficult due to the unequal particle masses. The kinetic energy of the system is initially damped (to around $10^{-6} E_{tot}$) but climbs at later times (to around $10^{-5} E_{tot}$) as the particle disorder spreads through the system. The particles at this stage appear to continually exchange between shifting blocks which do not seem able to settle to a minimum energy state (appearing not unlike the movement of shifting ice-floes).
  Our understanding of this phenomena is that the square lattice configuration with $\eta = 1.2$ in (\ref{eq:hrho}) is a quasi-stable state for the particles. This can be shown by a straightforward stability analysis in two dimensions \citep{morrisphd,bot04} which shows that the square lattice is unstable at certain ratios of the smoothing length to the particle spacing (although stable at $\eta = 1.2$). Given enough sensitivity to their configuration (through the smoothing length iterations) the particles will shift into a more isotropic arrangement similar to that observed in the equal mass case, however in this case a minimum energy configuration is more difficult to obtain with unequal particle masses, since the masses have been set to correspond to the equilibrium density profile using the initial particle configuration.

 These results show that it is preferable to place the unequal mass particles initially on a lattice of hexagonal cells since this is the configuration to which they tend to evolve.

  We therefore set up the particles in this manner, with the (hexagonal) lattice centred on the origin and again with the particle masses set according to the equilibrium density profile. 1045 particles are used.  In this case the lattice is able to damp to an equilibrium state since the particles do not shift from the lattice apart from a small initial adjustment at the outer edges (note that we also use a $\rho_{min}$ in (\ref{eq:hrho}) as will be discussed in \S\ref{sec:axisym}). The equilibrium distribution in this case is shown in Figure~\ref{fig:static_varm} and the density profile is shown in Figure~\ref{fig:static_varm_rho}. 

\subsubsection{Reversible time integration}
   The relaxation procedure for the Toy Star also revealed some pertinent aspects of the reversible integration algorithm. In particular if the timestep changes rapidly between steps, the energy in the reversible integrator could exhibit large oscillations and under some circumstances become unstable. This is a result of the averaging procedure used to determine the timestep. In this paper we have used the arithmetic average
\begin{equation}
\delta t^{1/2} = \frac12 (\delta t^{0} + \delta t^{1}),
\end{equation}
where the superscripts $0,\frac12$ and $1$ refer to the stepsize calculated at the beginning, middle and end of a given timestep. This expression is rearranged in order to determine the next timestep $\delta t^{1}$ from the previous timestep $\delta t^{0}$ and the timestep determined from the Courant condition at the half step $\delta t^{1/2}$. The pitfall of this method is that if the timestep changes rapidly (for example, from a very short step to a very long step) then the timestep begins to oscillate between two values (short-long-short-long). Also there is nothing to prevent the timestep $\delta t^{1}$ from taking a negative value. In this situation the time advance is achieved by taking a long step forward and a slightly shorter step back to achieve a small net forward step.
Whilst the specific problem of negative timesteps can be fixed by using a different averaging procedure (for example using the geometric mean), we have found that the oscillations can also cause instability if $\delta t^{1}$ greatly exceeds the Courant condition. This can occur because a very short $\delta t^{0}$ implies a very long $\delta t^{1}$ (where by definition $\delta t^{1/2}$ meets the Courant condition). If the Courant constraint also changes between steps the result can be that particles are evolved on a timestep which is too large to capture the change in physical quantities accurately, resulting in instability.

 In practice these oscillations can be largely avoided provided that some care is taken in the setup and evolution. For example in the initial runs we began the integration as we had done for the predictor-corrector method using a timestep of zero (which simply runs through the force evaluation without evolving the particles). In the reversible integrator this immediately leads to timestep oscillations since the zero timestep is then used in the averaging procedure. A second example is that it is common practice to reduce the timestep before output dumps in order to reach the specified time of output exactly. This means taking a shorter step just before an output dump but for the reversible integrator this will again trigger the timestep oscillations. Thus an interpolation (rather than an evolution) must be used to output the data at a specific time. Finally, in order to prevent any potential instability we place a check in the timestepping algorithm which resynchronises $\delta t^{0}$, $\delta t^{1/2}$ and $\delta t^{1}$ to the current value of $\delta t^{1/2}$ if $\delta t^{1}$ exceeds the Courant condition (but without a safety factor). This means using a reasonably low safety factor for the Courant condition used to find $\delta t^{1/2}$, in order to allow sufficient variation between $\delta t^{0}$, $\delta t^{1/2}$ and $\delta t^{1}$. Obviously any timestep resychronisation means that the evolution is no longer globally reversible, although the sections of the evolution between resynchronisations will be reversible.
 
 It should be noted that methods do now exist for reversible integration of SPH-type systems which do not suffer from timestep oscillation instabilities. These methods involve a rescaling of the time variable so that the timestep is a continuously variable quantity (Leimkuhler, private communication). The application of such methods to SPH will be investigated elsewhere.
\begin{figure*}
\begin{center}
\begin{turn}{270}\epsfig{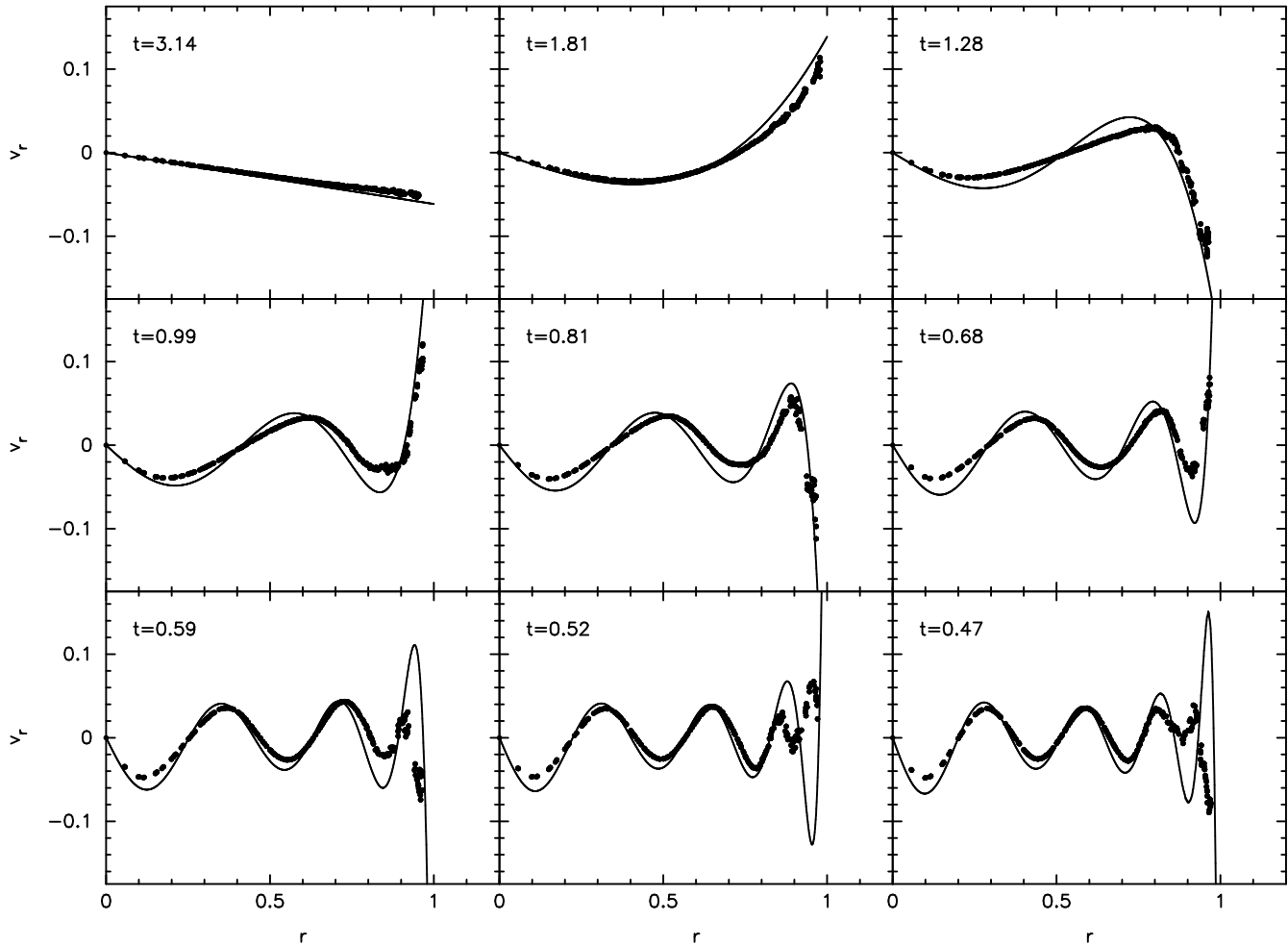}\end{turn}
\caption{Axisymmetric (s=0) modes j=2,4,6,8,10,12,14,16 and 18 after 1 (theoretical) oscillation period using 1045 particles with masses initially varied to give the equilibrium density profile. The axisymmetric modes exhibit strong oscillations near the outer edges of the Toy Star which can be difficult to capture numerically.}
\label{fig:rmodes}
\end{center}
\end{figure*}

\begin{figure}
\begin{center}
\begin{turn}{270}\epsfig{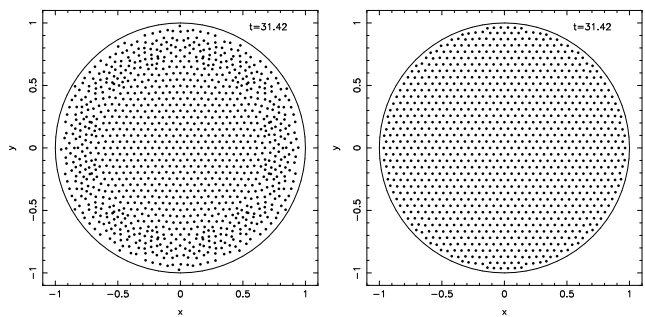}\end{turn}
\caption{Particle distribution in the unequal particle mass run after 10 oscillations of the axisymmetric $j=2$ mode. In the unequal mass case using $h=\eta (m/\rho)^{1/2}$ can lead to large smoothing length gradients at the edge of the Toy Star, causing some disruption at the outer edge (left figure). Using $h=\eta [m/(\rho + \rho_{min})]^{1/2}$ (right figure) prevents this problem from occurring.}
\label{fig:tvarm_mess}
\end{center}
\end{figure}

\begin{figure*}
\begin{center}
\begin{turn}{270}\epsfig{file=rmodes_deltarho.ps, width=0.44\textheight}\end{turn}
\caption{Density perturbations in the axisymmetric (s=0) modes j=2,4,6,8,10,12,14,16 and 18 using 10053 unequal mass particles. The difference between the density and the equilibrium density profile is plotted for each mode after 1.25 oscillation periods.}
\label{fig:rmodes_deltarho}

\begin{turn}{270}\epsfig{file=phimodes_deltarho.ps, width=0.44\textheight}\end{turn}
\caption{Density perturbations in the j=0 modes with s=2,4,6,8,10,12,14,16 and 18 using 10053 unequal mass particles. The difference between the density and the equilibrium density profile is plotted for each mode after 1.25 oscillation periods.}
\label{fig:phimodes_deltarho}
\end{center}
\end{figure*}

\begin{figure*}[h]
\begin{center}
\begin{turn}{270}\epsfig{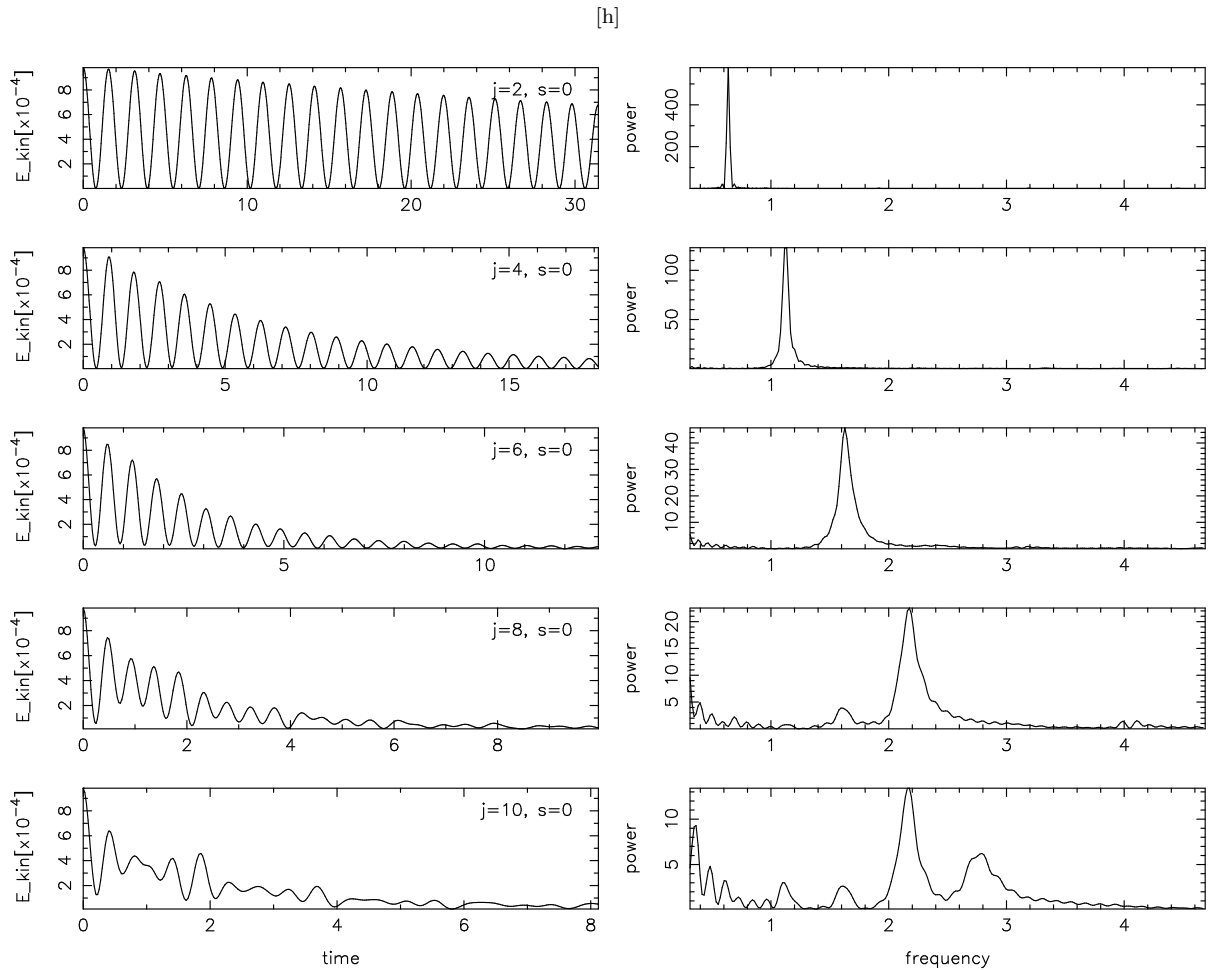}\end{turn}
\caption{Time evolution of the first few axisymmetric (s=0) modes (from top to bottom j=2,4,6,8 and 10) using 1000 equal mass particles. The left hand side shows the time evolution of the kinetic energy over 10 periods for each mode, whilst the right hand side shows the dominant frequencies of oscillation from the power spectrum (periodogram). From this plot we can see that the modes up to $j=8$ are captured well at this resolution (although the amplitudes are somewhat damped), whilst the higher order modes (in this case $j>10$) tend to decay into their lower order counterparts. Note that two oscillations of the kinetic energy correspond to one oscillation period.}
\label{fig:teqmlin_ev}
\end{center}
\end{figure*}
\subsection{Linear Oscillations}
 The linear oscillations of the two dimensional Toy Star may be examined by applying a velocity perturbation to the equilibrium configuration. We examine both the equal mass and unequal particle mass cases for both the axisymmetric and non-axisymmetric modes. The perturbation in each case is given to the velocity components according to (\ref{eq:vperturb}) (where the higher order modes are calculated using the recursion relation (\ref{eq:recurrencerelation})) and normalised such that the total kinetic energy 
\begin{equation}
E_{kin} = \frac12 (0.05 c_{s0})^{2},
\end{equation}
where $c_{s0}$ is the sound speed at the origin. 

The SPH simulations in each case were run for 10 periods of the given mode. The frequencies of oscillation in the numerical solutions may then be obtained from the kinetic energy evolution. The simplest method is to measure the spacing of minima and maxima in the kinetic energy. We use this method as well as calculating the full power spectrum of the kinetic energy evolution using the Lomb/Scargle periodogram for unevenly sampled data \citep{lomb76,scargle82}\citep[see][]{numericalrecipes}.

\subsubsection{Axisymmetric modes}
\label{sec:axisym}
 The linear, axisymmetric modes of the Toy Star are shown in Figure~\ref{fig:rmodes}, where the numerical solution in the unequal particle mass case is plotted for each mode after 1 oscillation period. The perturbation solution in each case is given by the solid line. The modes change rapidly near the boundary because the low density material respond to the push by getting a large velocity. This shows up in the rapid change of the hypergeometric functions which describe this variation. However, these strong oscillations can be difficult to capture numerically.  In the equal mass particle case at this resolution the modes are significantly damped by the artificial viscosity terms which act preferentially on the outer regions of the Toy Star\footnote{Note that the SPH artificial viscosity has a kinematic viscosity coefficient which has the form $\propto ( h C_s )$ which is constant near the edge since $h \propto 1/\sqrt{\rho}$ and $C_s \propto \sqrt{\rho}$ (when $P \propto \rho^2$). Whilst for other adiabatic equations of state the coefficient becomes infinite, in this case the damping is large in the outer regions purely because the second derivatives of the velocity are very large. }. In the unequal particle mass case the large fluctuations at the outer edges are better resolved since the resolution is constant throughout the Toy Star, however some problems occur at the edges due to mixing between particles of different masses caused by a combination of the large smoothing length gradient at the outer edge, the velocity fluctuations from the oscillations and the presence of very low mass particles there (Figure~\ref{fig:tvarm_mess}, left panel).
 
  The large smoothing length gradient at the edge of the Toy Star is largely a defect of our use of ($\ref{eq:hrho}$) in the case of unequal mass particles. Whilst the problem also occurs using standard SPH formalisms, it is accentuated by our use of the variable smoothing length formalism since the large gradients in the smoothing length are explicitly incorporated into the force terms. The relation ($\ref{eq:hrho}$) may cause problems in this case because, ideally, the smoothing length should relate to the particle number density (or roughly, the number of neighbours) rather than the mass density. In the case of equal mass particles ($\ref{eq:hrho}$) indeed has this effect. The problem in the unequal mass Toy Star is that where the smoothing length should increase at the outer radius (due to the decrease in particle number density), in practise the smoothing length in fact \emph{decreases} at the edge due to the drop-off in particle mass affecting the numerator in ($\ref{eq:hrho}$). A simple fix to this problem is to modify the relation using a density floor, giving
\begin{equation}
h = \eta \left(\frac{m}{\rho + \rho_{min}}\right)^{1/2},
\label{eq:hrhomin}
\end{equation}
where $\rho_{min} = {\rm min}(\rho_{prev})$, where $\rho_{prev}$ is the density on the particles from the previous timestep. This modification prevents the smoothing length gradients from becoming too large at the edge.

 Using a density floor is not particularly desirable in general. In fact it is quite simple to derive a generalisation of the variable smoothing length formalism of \citet{sh02} and \citet{monaghan02} in which the smoothing length is a function of the particle number density rather than the mass density\footnote{A similar formalism was described by \citet{os03} and derived self-consistently in \citet{price04} but has the disadvantage of also modifying the density evaluation which we have found to be somewhat problematic.}. However, since this is a paper on Toy Stars rather than on SPH formalisms the detailed description of this formalism will be presented elsewhere and we simply use (\ref{eq:hrhomin}) in this paper. In terms of the frequency estimate there is very little difference between the results using (\ref{eq:hrho}) and those using the density floor. In fact the disruption seen in (\ref{fig:tvarm_mess}) is largely cosmetic for short simulations since the very low mass particles do not have a strong influence on the overall evolution. For longer simulations the effect is more problematic as the disruption in the particle distribution spreads to the inner regions. It should also be noted however that these effects also become less significant as the total number of particles is increased.

The kinetic energy evolution in the first 5 axisymmetric modes in the case of equal mass particles is shown in Figure~\ref{fig:teqmlin_ev}. The left panel shows the kinetic energy evolution whilst the right panel shows the power spectrum of the left panel, showing the dominant frequency in each case. The modes show significant damping although the dominant frequency is captured well by the numerical solution up to the $j=10$ mode which can be seen to decay rapidly into the $j=8$ mode.
\begin{figure}
\begin{center}
\begin{turn}{270}\epsfig{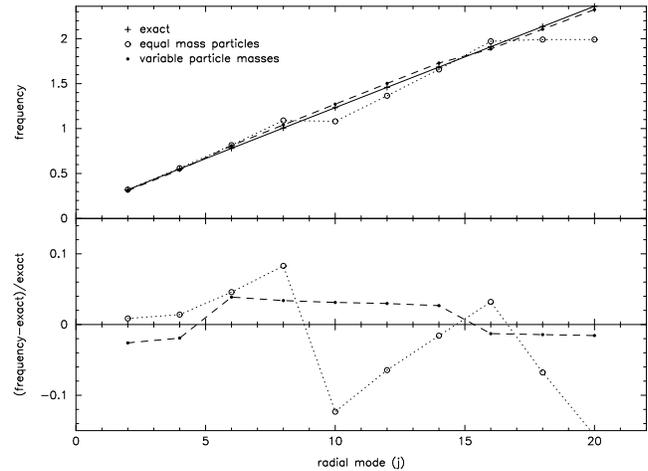}\end{turn}
\caption{Frequencies of the axisymmetric modes at a resolution of 1000 particles. The top panel shows the absolute frequencies whilst the bottom panel shows the fractional error in the numerical solution. The simulations using unequal particle masses, denoted by filled circles (a connecting line (dashed) has been plotted for clarity) are within $2$ percent of the correct frequencies for modes up to $j=20$. The results using equal mass particles (open circles, dotted line) show errors of up to $10\%$ for modes above $j=6$ although the general trend is still observed.}
\label{fig:rfreq}
\end{center}
\end{figure}

\begin{figure}
\begin{center}
\begin{turn}{270}\epsfig{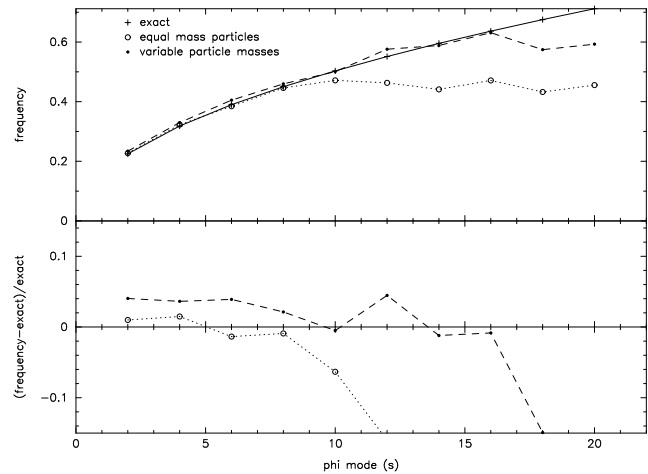}\end{turn}
\caption{Frequencies of the $j=0$ modes at a resolution of 1000 particles. The top panel shows the absolute frequencies whilst the bottom panel shows the fractional error in the numerical solution. The simulations using unequal particle masses, denoted by filled circles (where again a connecting line (dashed) has been plotted for clarity) are within $2$ percent of the correct frequencies for modes up to $s=16$. The results using equal mass particles (open circles, dotted line) show somewhat lower errors but only for modes up to $s=8$ above which the modes are largely damped out.}
\label{fig:phifreq}
\end{center}
\end{figure}

\begin{figure*}
\begin{center}
\begin{turn}{270}\epsfig{file=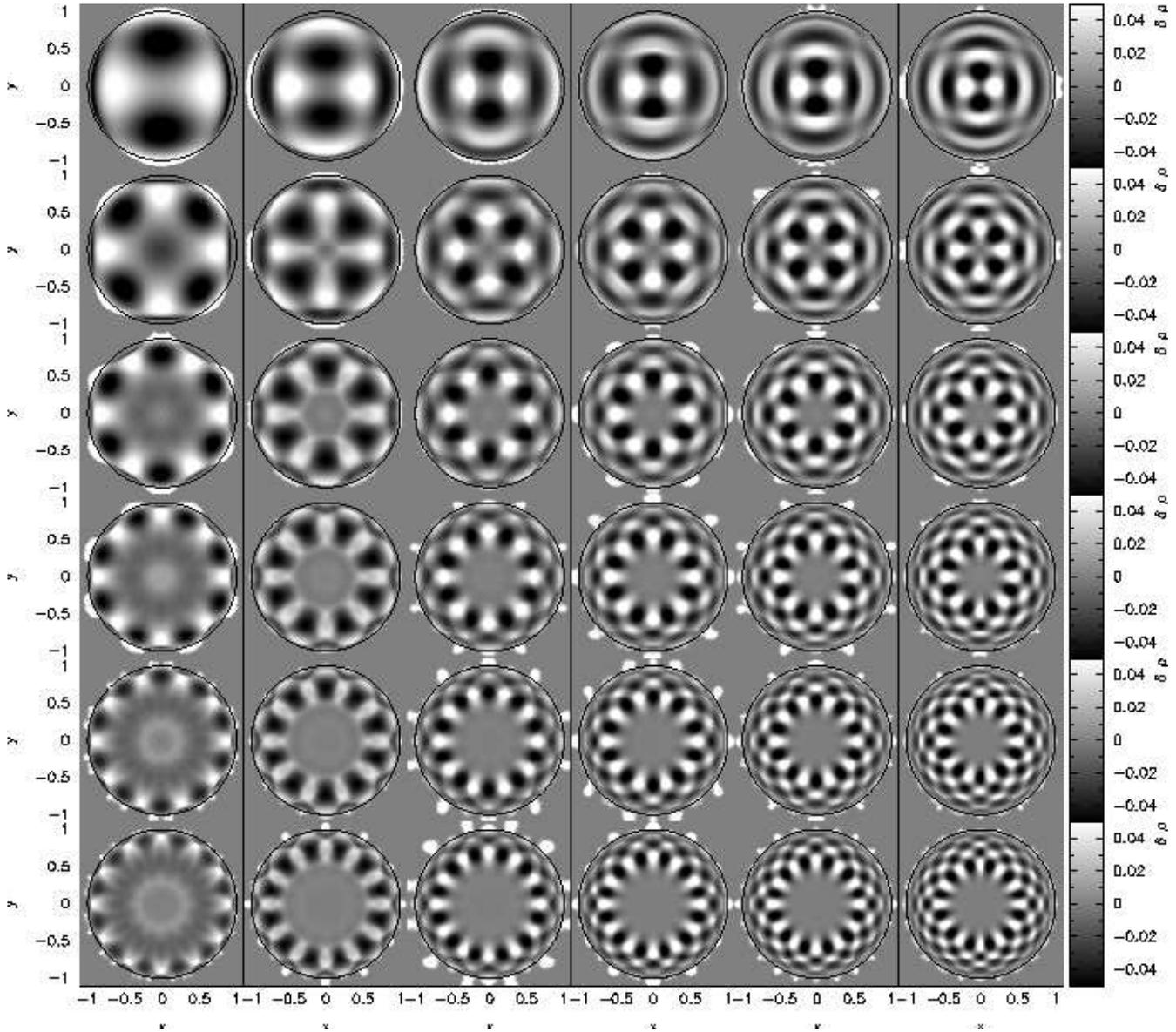,height=\textwidth}\end{turn}
\caption{Density perturbations for the mixed modes using 10053 unequal mass particles. From left to right the radial mode number varies from $j=2$ (leftmost column) to $j=12$ (rightmost column) whilst from top to bottom the angular mode number varies from $s=2$ (top row) to $s=12$ (bottom row). The difference between the density and the equilibrium density profile is plotted for each mode after 1.25 oscillation periods.}
\label{fig:mixedmodes_deltarho}
\end{center}
\end{figure*}

  The frequencies in the numerical solutions are shown in Figure~\ref{fig:rfreq}. The top panel shows the absolute frequency for each axisymmetric mode plotted as a function of the mode number $j$. The frequency plotted in each case is the dominant frequency measured from the power spectrum of the kinetic energy evolution for the simulation of that particular mode. The exact frequencies are given by the crosses, where a solid connecting line has been drawn for clarity. Results using equal mass particles are given by the open circles (with a dashed connecting line for clarity) whilst the unequal particle mass results are indicated by the filled circles (with a dotted connecting line). The lower panel shows the error in the numerical frequency (shown as a fractional deviation from the exact value) in each case. The unequal particle mass simulations at this resolution show errors of $\lesssim 2\%$ for all modes up to $j=18$. The results using equal mass particles have comparable accuracy up to the $j=6$ mode but thereafter show errors of $\sim 10\%$. One interesting point to note in the equal particle mass case is that the $j=10$ mode seems to decay to the $j=8$ mode and correspondingly the $j=20$ mode seems to decay to the $j=16$ mode. The frequencies obtained with the time-reversible integrator are indistinguishable from those obtained using the predictor-corrector method, provided that the timestep does not change rapidly between timesteps as discussed in \S\ref{sec:static}.

\subsection{Non-axisymmetric oscillations}
\subsubsection{$j=0$ modes}
 The two dimensional Toy Star can also oscillate non-axisymmetrically. The frequencies in the numerical solutions at a resolution of 1000 particles for the $j=0$ modes (j=0, s=2,4,6 etc) are shown in Figure~\ref{fig:phifreq}. As in Figure~\ref{fig:rfreq} the absolute frequencies are shown in the top panel whilst the fractional error is shown in the bottom panel. Using unequal mass particles (filled circles, dashed line) the frequencies in the numerical solutions show good agreement with the exact solutions (errors $\lesssim 2\%$) for modes up to $s=16$, above which the modes are damped out. The results using equal mass particles (open circles, dotted line) show somewhat lower errors ($\lesssim 0.6\%$) but only the modes up to $s=8$ are captured in this case.

\subsubsection{Mixed modes}
 Finally the Toy Star can oscillate with modes which have both $j\neq 0$ and $s\neq 0$. The density perturbation in the modes with $2\le j \le 12$ and $2\le s \le 12$ is shown in Figure~\ref{fig:mixedmodes_deltarho}. The numerical frequencies of the perturbations show similar accuracy to the symmetric modes. 


\section{Exact, non-linear solutions}

As mentioned earlier, one of the most useful features of the Toy Star is that it is possible to get exact, or very accurate solutions of the full non linear equations.  These provide excellent tests of the ability of a numerical code to follow a surface into a region devoid of matter.  We assume that the velocity field is a linear function of the cartesian coordinates

\begin{eqnarray}
v_x & =  &V^{11} x + V^{12}y, \label{eq:vxnonaxi} \\
v_y  & = &V^{21}x + V^{22}y, \label{eq:vynonaxi}
\end{eqnarray}
and that the density is given by the following second degree function of the coordinates
\begin{equation}
\rho^{\gamma-1} = H(t) - [ C(t) x^2 + 2 B(t) xy + D(t) y^2]. \label{eq:rhononaxi}
\end{equation}

\subsection{Axisymmetric nonlinear solutions}
For this case we choose $V^{11} = V^{22} = \alpha(t)$ and $V^{12} = - V^{21} = - \beta(t)$ in which case the velocity field can be written
\begin{equation}
{\bf v} = \alpha(t) {\bf r} + \beta(t) \hat{ \bf {z} } \times {\bf r},
\label{eq:vnonlin}
\end{equation}
where the first term produces an axisymmetric expansion or contraction, and the second term gives a rigid rotation with angular velocity $\beta(t) \hat{\bf z}$ where $\hat{\bf z}$ is a unit vector in the z direction. The motion takes place so that, at any time, each element of fluid has the same angular velocity.   The reader will appreciate that the solution has many features in common with the symmetric pulsation of a polytropic star, for example a model White Dwarf, under gravity. 

For this  axisymmetric case  $C=D$,   $B=0$,  so that
\begin{equation}
\rho^{\gamma-1} = H(t) - C(t) r^2.
\label{eq:rhoHC}
\end{equation}

Substitution of the assumed forms for ${\bf v}$ and $\rho$ into the acceleration and continuity equations, with $P = K \rho^\gamma$ and replacing the Lagrangian derivative  by the Eulerian according to
\begin{equation}
\frac{d}{dt} = \frac{\partial}{\partial t} + (\bf {v} \cdot \nabla) ,
\end{equation}
 and equating powers of $x$ and $y$ we get

\begin{eqnarray}
\dot{\alpha } & = & -\alpha^2 + \beta^2 + \frac{2CK \gamma}{\gamma-1} - \Omega^2, \label{eq:alphadot}\\
\dot{\beta} & = & -2 \alpha \beta, \label{eq:betadot} \\
\dot{H}&   = &-2 \alpha (\gamma-1) H,\\
\dot{C}& = & -2 C \alpha \gamma. \label{eq:Cdot} 
\end{eqnarray}
Where the $\dot{a}$ denotes the time derivative  of  any  function $a$. 

If $\gamma = 2$, a further derivative of (\ref{eq:alphadot}) with respect to time using (\ref{eq:betadot}) to remove the derivative of $\beta$ results in the following equation for $\alpha$
\begin{equation}
\ddot{\alpha}+ 6 \alpha \dot{\alpha} + 4 \alpha^3 + 4 \alpha \Omega^2 = 0.
\end{equation}
If $\alpha$ is sufficiently small this equation can be approximated by
\begin{equation}
\ddot{\alpha}+ 4 \alpha \Omega = 0. \label{eq:alphalinearsol}
\end{equation}
with solution $\alpha = A \sin{(2 \Omega t + c)}$ where $c$ is an arbitrary constant.   An exact solution of the full non linear  equation can also be found.  The analysis is given in Appendix \ref{sec:alphaexactsolution}. 

The set of ordinary differential equations (\ref{eq:alphadot})-(\ref{eq:Cdot}) can be integrated with high accuracy using any standard stable ODE integrator.

\subsubsection{ Conserved quantities}

The total mass is
\begin{equation}
 M =  \frac{ \pi H ^{ \frac{\gamma}{\gamma-1} }}{C}  \frac{ (\gamma-1) }{\gamma},
\end{equation}
and must be conserved by the set of equations  (\ref{eq:alphadot})-(\ref{eq:Cdot}).  To see that this is the case we eliminate $\alpha$ from the last two of these equations to get
\begin{equation}
\frac{dH}{dC} = \frac{ (\gamma-1)H }{\gamma C},
\end{equation}
from which  $H^\gamma \propto C^{\gamma-1}$ showing that $M$ is constant.  The angular momentum $J_z$ should also be conserved by the equations.   $J_z$ is given by 
\begin{equation}
J_z = \int \int \rho (x v_x - y v_y) dx dy = 2 \pi \beta  \int  [H-Cr^2]^\frac{1}{\gamma-1} r^3 dr.
\end{equation}
Completing the integration shows that 
\begin{equation}
J_z = 2 \pi \beta \frac{H^{\frac {2 \gamma-1}{\gamma-1} } }{C^2}\frac{(\gamma-1)^2}{2 \gamma (2 \gamma-1)}.
\label{eq:Jz}
\end{equation}
But from (\ref{eq:betadot}) and (\ref{eq:Cdot}) we find $\beta^\gamma \propto C $. Substitution into (\ref{eq:Jz}) and noting that  $H^\gamma \propto C^{\gamma-1}$  we find $J_z$ is constant. Of course the system must conserve $M$ and $J_z$ and the foregoing analysis merely confirms that the equations we have derived are correct.

\subsection{The Lagrangian description}
In this section we consider the solutions in terms of the change in the functions $H$, $C$ and $\beta$ {\it for a given element } of fluid.  Defining the function $S(t) $ by
\begin{equation}
S(t) = exp{ (- \int _0^t  \alpha dt)}.
\end{equation}
we can write
\begin{eqnarray}
\beta  & = & \beta_0 S^2 ,\\
H& = & H_0 S^{2(\gamma-1) }, \\
C& = & C_0 S^{2 \gamma},
\end{eqnarray}
and from the continuity equation (\ref{eq:cty})
\begin{equation}
\rho = \rho_0 S^{2}
\label{eq:rhos2}
\end{equation}
where we assume the motion begins at $t=0$.  These equations can be interpreted in the following Lagrangian way.  For any element of fluid (particle), the initial value of one of these functions, say $\beta$, is denoted by $\beta_0$. The equation for $\beta$  in terms of $S$ gives its value for that particle at any time.  Similarly for the other quantities.  This also enables one quantity to be written  in terms of the others. For example we deduce $H \beta \propto C$.  The  $x,y$ coordinates of a particle can be obtained from  the velocity according to
\begin{eqnarray}
\frac{dx}{dt}& = & \alpha x - \beta y, \\
\frac{dy}{dt} & = & \alpha y + \beta x ,
\end{eqnarray}
  From these equations we easily deduce that the radial coordinate $r$ and angular coordinate $\theta$ of the particle are given by
\begin{eqnarray}
r^2 & = & r^2_0 S^{-2}, \\
\theta & = & \int_0^t \beta(t) dt,
\end{eqnarray}
from which we deduce that for any  particle $Cr^2 \propto  H$ as is required by the density equation (\ref{eq:rhoHC}) and (\ref{eq:rhos2}). 

A further useful result is that equations (\ref{eq:alphadot}) and (\ref{eq:betadot}) can be combined together with the  relation $C = \sigma \beta^\gamma $,  to give
\begin{equation}
\frac{d \alpha}{d \beta} =   \frac{1}{2  \alpha \beta} \left  (  \alpha^2  + \Omega^2 - \beta^2- \frac{2 \gamma K}{\gamma -1} \sigma \beta^\gamma     \right ),
\end{equation}
This equation can be integrated exactly taking $\alpha^2$ as a new dependent variable to give a linear differential equation.  In this way we find 
\begin{equation}
\alpha^2 + \beta^2 +  \Omega^2 =  - \frac{2 K \gamma C}{(\gamma-1)^2 } +  k \beta,
\label{eq:alphabeta}
\end{equation}
where $k$ is an integration constant.  This equation can also be derived from the constancy of energy. (\ref{eq:alphabeta}) defines a closed curve in the $\alpha$ $\beta$ plane showing that the motion is a periodic oscillation (demonstrated in Figure~\ref{fig:alphabeta}).

\subsection{ SPH simulations of the  non linear  axisymmetric motion}
 The non-linear axisymmetric Toy Star is set up by perturbing the equilibrium solution with a velocity of the form (\ref{eq:vnonlin}) where the parameters $\alpha(0)$ and $\beta(0)$ specify the amplitudes of the initial expansion and rotation respectively.

\begin{figure}
\begin{center}
\begin{turn}{270}\epsfig{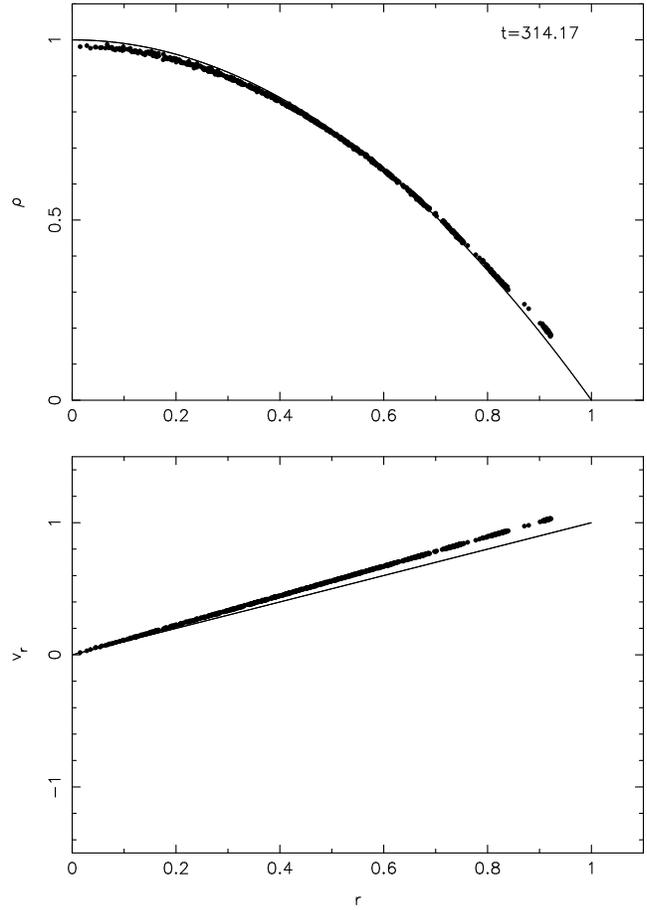}\end{turn}
\caption{Density and radial velocity in the non-linear Toy Star solution after 100 oscillation periods. The SPH particles are indicated by the solid points whilst the exact solution is given by the solid line. }
\label{fig:densvrnonlin}
\end{center}
\end{figure}

 Since the axisymmetric mode given by (\ref{eq:vnonlin}) is a non-linear, exact solution the choice of amplitudes is completely arbitrary. We chose for simplicity, $\alpha_{0} = 1$ and $\beta_{0} = \pi$. This means that the expansion velocity is everywhere supersonic (ie. highly non-linear) and that initially the rotation rate was equal to the oscillation period (note that the actual rotation period during the evolution depends on the periodic manner in which $\beta$ varies). The SPH simulation was run by perturbing the equal mass particle equilibrium configuration. No artificial viscosity was applied during the evolution. The density and radial velocity are shown in Figure~\ref{fig:densvrnonlin} after 100 oscillation periods. The agreement with the exact solution after 100 periods is extremely good, with both the density and velocity profiles maintained almost exactly. 

\begin{figure}
\begin{center}
\begin{turn}{270}\epsfig{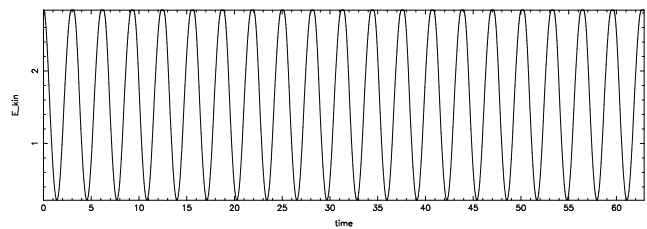}\end{turn}
\caption{Time evolution of the nonlinear axisymmetric mode of the Toy Star using 1000 equal mass particles, where initially $\alpha = 1$ and $\beta = \pi$. For clarity only the first 20 oscillation periods are plotted. The amplitude is maintained exactly by the SPH solution for more than 1000 periods.}
\label{fig:tnonlin_ev}
\end{center}
\end{figure}

 The time evolution of the total kinetic energy in the simulation is shown in Figure~\ref{fig:tnonlin_ev}, showing only the first 20 periods for clarity. The amplitude of the oscillation is maintained exactly by the SPH solution.

\begin{figure}
\begin{center}
\begin{turn}{270}\epsfig{file=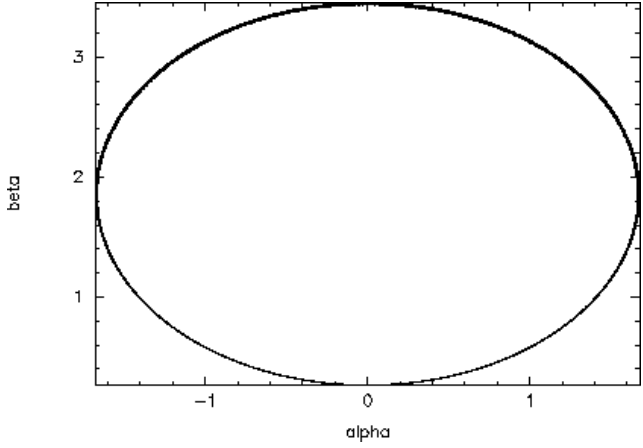, height=\columnwidth}\end{turn}
\caption{The Lagrangian quantities $\alpha$ and $\beta$ measured from the SPH solution using 1000 equal mass particles. The average values of $\alpha$ and $\beta$ on the particles has been plotted every timestep for over 1000 oscillation periods (t=3200, 72,863 timesteps).}
\label{fig:alphabeta}
\end{center}
\end{figure}

 Finally it is useful to plot the evolution of the Lagrangian quantities $\alpha$ and $\beta$ in the numerical solution. In order to do so we calculate $\alpha_{i} = ({\bf v}\cdot{\bf \hat{r}})_{i}$ and $\beta_{i} = ({\bf \hat{r}}\times {\bf v})_{z,i}$ for each particle $i$ after every timestep and then compute the average over all the particles. These average values of $\alpha$ and $\beta$ are plotted in Figure~\ref{fig:alphabeta}, where each point in the plot corresponds to the values at a given timestep. The curve agrees exactly with the analytic solution given by (\ref{eq:alphabeta}).

 The results of this simulation demonstrate that SPH has extremely good conservation properties and can maintain an accurate evolution of a non-linear system over long time intervals. We have experimented with a range of values for $\alpha_{0}$ and $\beta_{0}$ all of which show similar results. In practise the necessary addition of artificial viscosity into SPH algorithms in order to handle shocks leads to unphysical damping of self-similar motion such as that which occurs in the non-linear Toy Star. This test would therefore be an excellent benchmark for switches designed to turn off the artificial viscosity away from shocks where it is not needed.

\subsection{Non-axisymmetric solutions}

 If the expressions (\ref{eq:vxnonaxi})-(\ref{eq:rhononaxi})  for the velocity and $\rho^{1/(\gamma-1)}$ are substituted into the equations of motion and the coefficients of powers of $x$ and $y$ and $xy$ the following equations are obtained from the acceleration equation 
\begin{eqnarray}
\frac{dV^{11}} {dt} & = & \frac{2 \gamma K}{\gamma-1} C - (V^{11} )^2 - V^{12} V^{21} - \Omega^2, \label{eq:dV11dt}\\
\frac{dV^{22} }{dt} & = &  \frac{2 \gamma K}{\gamma-1} D - (V^{22} )^2 - V^{12} V^{21} - \Omega^2,\\
\frac{dV^{12} }{dt} & = &  \frac{2 \gamma K}{\gamma-1} B - V^{12}( V^{11} + V^{22} ), \\
\frac{dV^{21} }{dt} & = &  \frac{2 \gamma K}{\gamma-1} B - V^{21}( V^{11} + V^{22} ),
\end{eqnarray}
and from the continuity equation we get
\begin{eqnarray} 
\frac{dH}{dt} &= & -(\gamma-1)(V^{11} + V^{22} ) H, \label{eq:dHdt}\\
\frac{dC}{dt}& = &- 2CV^{11} - (\gamma-1) C (V^{11} + V^{22} ) - 2B V^{21}, \\
\frac{dD}{dt} &=& - 2DV^{22} - (\gamma-1) D(V^{11} + V^{22} ) - 2B V^{12}, \\
\frac{dB}{dt} &= &-C V^{12} - DV^{21}  - \gamma B (V^{11} + V^{22} ) \label{eq:dBdtnonaxi}.
\end{eqnarray}
These 8 ordinary differential equations describe the system.  They can be integrated as accurately as desired using standard methods for differential equations.

\begin{figure*}
\begin{center}
\epsfig{file=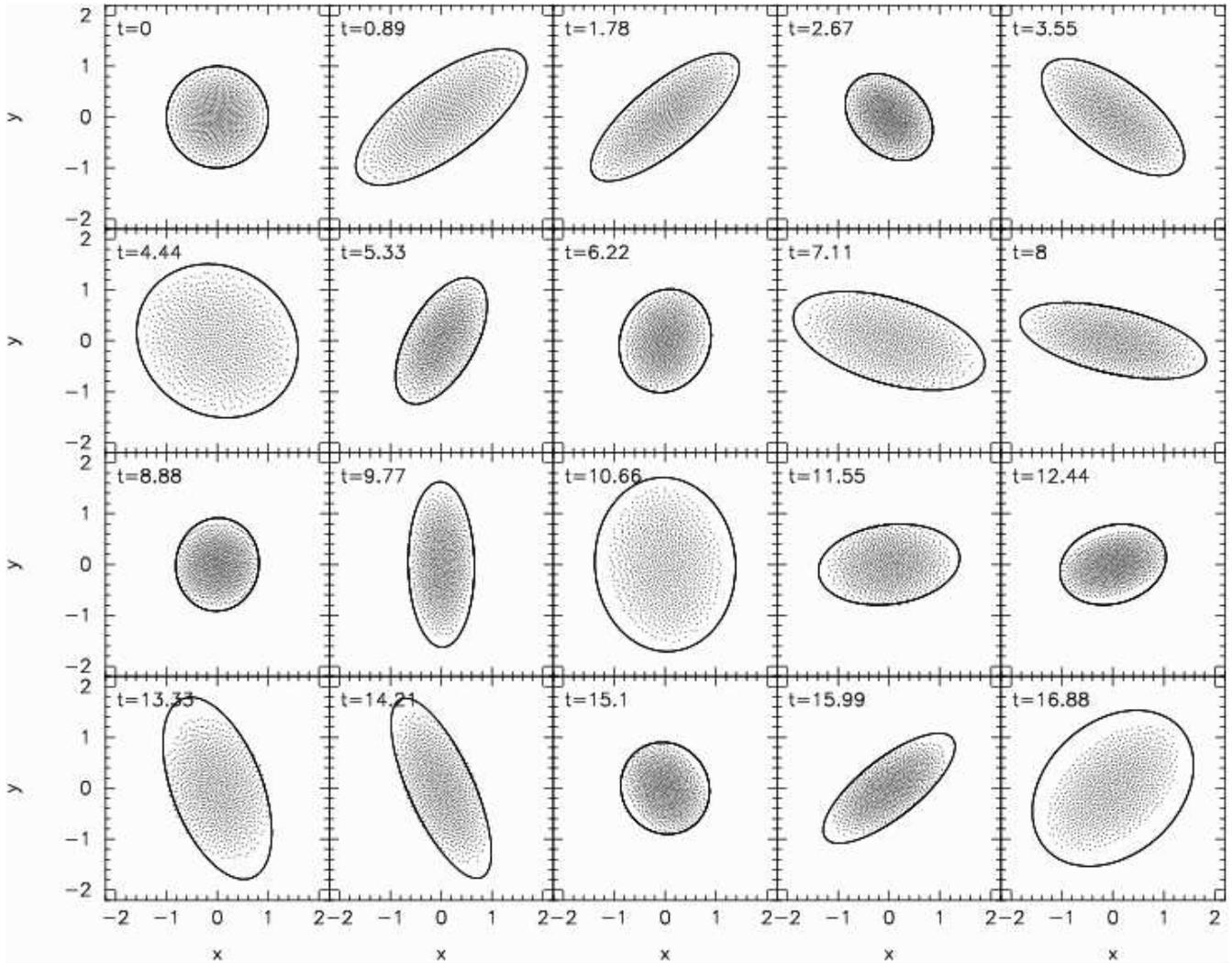, width=\textwidth}
\caption{Particle distribution during the evolution of the nonlinear, non-axisymmetric mode of the Toy Star using 1000 equal mass particles, where initially $V_{11} = 1$, $V_{12}=1/2$, $V_{21}=1$ and $V_{22} = 1/4$. A small damping of the amplitude may be observed in the SPH solution due to the artificial viscosity applied to prevent particle interpenetration during the compression phase.}
\label{fig:teqmnonaxi}
\end{center}
\end{figure*}

\subsubsection{Conserved Quantities}
The mass is given by 
\begin{equation}
M =\int \int ( H - (Cx^2 + 2Bxy + Dy^2) )^{1/(\gamma-1) }dx dy 
\end{equation}
This integral can be evaluated by rotating the coordinate system so that the quadratic form in the integrand is reduced to a sum of squares.  The final result is
\begin{equation}
M = \pi \left ( \frac{\gamma-1}{\gamma} \right ) \frac{H^{\gamma/(\gamma-1) } } {\sqrt{ CD -B^2 }}.
\label{eq:massnonaxi}
\end{equation}
From the differential equations we find 
\begin{equation}
\frac{ d ( CD -B^2) }{dt} = - 2 \gamma (CD-B^2) ( V^{11} + V^{22} ), \label{eq:dCDmBdt}
\end{equation}
which, combined with (\ref{eq:dHdt}) shows that $M$ is conserved. 

The angular momentum $J_z$ is given by 
\begin{equation}
J_z =\int \int ( H - (Cx^2 + 2Bxy + Dy^2))^{1/(\gamma-1)} X dx dy,
\end{equation}
where we have defined
\begin{equation}
X \equiv \left ( x^2 V^{21} + xy( V^{22} -V^{11}) - y^2 V^{12}   \right ).
\end{equation}
Rotating the coordinate system as before the integral reduces to
\begin{equation}
J_z =  \frac{\pi (\gamma-1)^2}{ \gamma (2 \gamma -1) } \frac {H^{(2 \gamma -1)/(\gamma-1) } }{ (CD-B^2)^{3/2} } \Upsilon,
\end{equation}
where we define
\begin{equation}
\Upsilon \equiv V^{21}D-V^{12}C - B(V^{22} - V^{11}).
\end{equation}
From the differential equations above it is possible to show that the quantity $\Upsilon$
satisfies the equation
\begin{equation}
\frac{ d \Upsilon}{dt} = -(\gamma +1 )\Upsilon (V^{11}+V^{22} )
\end{equation}
In terms of the variable $T$ defined by
\begin{equation}
T = exp{\left  (- \int_0^t (V^{11} + V^{22} ) dt \right ) }
\end{equation}
we can write from (\ref{eq:dHdt}), (\ref{eq:dCDmBdt}) and the previous equation
\begin{eqnarray}
H& = & H_0 T^{(\gamma -1)}, \\
(CD -B^2) & = & (CD-B^2)_0 T^{2 \gamma} , \\
\Upsilon&  = &\Upsilon _0 T^{\gamma +1 }.
\end{eqnarray}
When these results are substituted into the previous expression for $J_z$ it is found to be constant, as expected.  As in the axisymmetric case proving the constancy of $M$ and $J_z$ from our set of differential equations merely confirms that they are correct.

Another conserved quantity is the circulation.  The circulation around any loop for this two dimensional fluid is given by
\begin{equation}
  \mathcal{C} = \oint {\bf v} \cdot {\bf dl} = \int \int \nabla \times {\bf v} dx dy.
\end{equation}
This is a constant of the motion so that its time derivative 
\begin{equation}\frac{ d \mathcal{c} }{dt} = \frac{d}{dt} \int \int \nabla \times {\bf v} dx dy = \int \int \frac{d}{dt} \left ( \frac{ \nabla \times {\bf v} }{\rho} \right ) \rho dx dy,
\end{equation}  
must vanish.  Noting that $\nabla \times {\bf v} = (V^{21} - V^{12} ) \hat{ {\bf z} } $ and that from the equations of motion
\begin{equation}
V^{21} - V^{12}  = (V^{21} - V^{12} )_0 T,
\end{equation}
and 
\begin{equation}
\rho  = \rho_0 T,
\end{equation} 
we see that $(\nabla \times {\bf v})/ \rho$ is unchanged following the motion and $\mathcal{C}$  is conserved.


\subsection{ SPH simulations of the non linear non axisymmetric motion}
 The non-axisymmetric exact solutions present a more demanding test problem since the Toy Star changes shape throughout the oscillation. This changing shape must then be followed through the expansion and contraction phases, maintaining the free boundary. The non-axisymmetric Toy Star therefore presents an ideal problem for numerical codes designed to simulate astrophysical systems with changing geometries and free boundaries.
 
  The simulation is set up by again perturbing the equilibrium solution, in this case with a velocity of the form (\ref{eq:vxnonaxi})-(\ref{eq:vynonaxi}) The four parameters $V_{11}, V_{12}, V_{21}$ and $V_{22}$ specify the amplitude and geometry of the perturbation. Due to the changing shape the SPH solution in this case requires some artificial viscosity in order to prevent particle interpenetration during the compression phase. We apply this using the \citet{mm97} switch as discussed in \S\ref{sec:sph}. 
 
  The particle distribution during the evolution of the exact, non-axisymmetric mode is shown in Figure~\ref{fig:teqmnonaxi}, where initially $V_{11} = 1$, $V_{12}=1/2$, $V_{21}=1$ and $V_{22} = 1/4$. For comparison with the exact solution we solve (\ref{eq:dV11dt})-(\ref{eq:dBdtnonaxi}) using a simple second order modified Euler predictor-corrector method, using the conservation of mass (\ref{eq:massnonaxi}) as a check on the quality of the integration. The solid line shown in Figure~\ref{fig:teqmnonaxi} is the curve corresponding to the edge of the Toy Star (ie. where $\rho=0$) at the appropriate times. The SPH particles adjust to the changing shape quite well apart from a damping of the amplitude with time caused by the application of artificial viscosity. We have performed a range of simulations using different values of the initial parameters which in general show very similar results.
  
   For simulations with very strong compression the particles can clump together in the manner described in \S\ref{sec:static} due to the force being zero at the origin of the cubic spline kernel used in the calculations, despite using $h=1.2 (m/\rho)^{1/2}$. This is because a strong compression can push the particles neighbours close enough to be in the region of the kernel where the force decreases towards the origin, causing a clumping instability for positive pressures in compression similar in its effect to the well known instability for negative stresses in tension \citep{sha95}. As discussed in \S\ref{sec:static} the remedy for this is to use a kernel with non-zero derivative at the origin, an investigation of which will be performed elsewhere.


\section{Summary and Discussion}
The primary aim of this paper has been to provide a set of benchmarks for simulations of gaseous disks of the kind that arise in star formation. The systems, which we call Toy Stars, are similar to their astrophysical counterparts in that  they consist of compressible gas, held together by an attractive force which results in the gas having an outer surface where the density and pressure fall to zero.  They therefore provide tests which are quite different to the usual tests based on flow in periodic rectangular regions. Furthermore, not only can the linear modes be found in terms of known functions, but non linear solutions can be found in terms of a small number of differential equations.

 The results described and discussed in this paper can be summarized as follows:

\begin{enumerate}
\item The static structure has been simulated using SPH with a variety of initial configurations using both equal mass and unequal mass particles.  The calculations use approximately 1000 particles. With this number of particles the agreement of the static model with theory is very good.  The final arrangement of the particles is different depending on whether they are equal mass or unequal mass. Recommendations concerning damping the particles to equilibrium are contained in the text. 
 
\item  The SPH simulations use the equations which result from using a variational principles and taking the resolution length to be a function of the density. The form of this relation, and effects that occur for different choices of parameters is discussed in detail.
 
\item The theoretical  linear modes of oscillation have been derived and they have been calculated using SPH.  Because the resolution of the SPH calculations is not high  the simulation would not be expected to recover modes high order modes. In fact, with 1000 particles the particle spacing is typically 0.06 (though larger near the edge  in the case of equal mass particles). Accordingly, since several particle spacing between nodes is normally required to give reasonable accuracy, the linear modes of the Toy Star with mode number greater than about 3 would not be expected to be accurate. Remarkably the SPH simulations give good frequencies for radial and azimuthal mode numbers ($s$) much larger than this. The modes, however, are not simulated as accurately. We note, in particular, that in the outer layers of the Toy Stars where the density and speed of sound falls to zero, the modes vary rapidly, and they would only be reproduced accurately by a simulation with many more particles (as for example in Figures~\ref{fig:rmodes_deltarho}, \ref{fig:phimodes_deltarho} and \ref{fig:mixedmodes_deltarho} used to illustrate the higher order modes).
 
\item Our analysis of a class of non-linear modes reduces the partial differential equations to a set of 8 non linear ordinary differential equations. These have been derived, with a different interpretation, in the context of the oscillation of water in basins.  The equations describe a rotating oscillating ellipse of gas which is similar to what would be expected when the attractive force is gravity.  The SPH simulation of this system is in good agreement with the accurate integration of the ordinary differential equations.
 
\item Our results show that SPH simulations of compact rotating masses of gas are very satisfactory. The actual accuracy depends, of course, on the resolution and therefore the number of particles used. 
 
\end{enumerate}


\section*{Acknowledgments}
  DJP is supported by a PPARC postdoctoral research fellowship. He also acknowledges the support of the Commonwealth Scholarship Commission and the Cambridge Commonwealth Trust. DJP also wishes to thank Monash University for their hospitality during a visit.

\appendix
\section{Linear Modes}
\label{sec:linearmodeappendix}
\subsection{Axisymmetric modes}

For the axisymmetric modes $s=0$.  The lowest axisymmetric mode has $j=2$ and setting 
\begin{equation}
 a_0 = f \rho_0^{\gamma -1}
 \end{equation}
we find
\begin{eqnarray}
\zeta& = &f \rho_o^{\gamma-1}  \left ( 1 -  \frac{\gamma}{\gamma-1} X^2 \right ),\\
{\bf V} &= &-f \left (   \frac{\Omega^2}{ \sigma (\gamma-1}  \right ) {\bf r}
\end{eqnarray}
For the next mode ($j=4$)
\begin{eqnarray}
\zeta& =& f \rho_0^{\gamma-1} \left ( 1 -  \frac14  \nu_4^2 X^2 + \frac{\nu_4^2 (3 \gamma -2)}{16(\gamma-1)} X^4 \right ),\\
{\bf V}& = & -f \frac{\Omega^2 \nu_4^2}{ 4 \sigma } \left ( 1 - \frac{3 \gamma-2}{2 (\gamma-1) } \frac{r^2}{r_e^2}  \right  ){\bf r}.
\end{eqnarray}
where $\nu_4^2 = 4(4 \gamma-2)/(\gamma -1)$.  Higher order terms can be obtained easily using the recurrence relation for the $a_k$. 

\subsection{Non-axisymmetric modes}

The non axisymmetric case $s=1$ is shift of the system.  To see this in the simplest case we note that the lowest mode has $j=0$ and the spatial part of the  perturbation $\eta$ for the choice $\sin{\theta}$ for the angular variation, has the form
\begin{equation}
D = a_0 r \sin{\theta} = a_o x,
\end{equation}
so that the $x$ component of the centre of mass becomes
\begin{equation}
\int \int  \left (\bar{ \rho} + \frac{( \bar{\rho})^{2 - \gamma} } {\gamma-1} \eta \right ) x rdr d \theta.
\end{equation}
When $s=1$, $D$ is an odd function in $x$ so the $x$ component of the centre of mass is non zero.  This mode therefore corresponds to the centre of mass shifting back and forth in the $x$ direction which is impossible for an isolated object.  On the other hand, it is entirely possible in a lake of water perturbed by tidal forces or a glass of wine moved back and forth.  It can be shown that, this mode ($s=1$ and $j=0$ ) results in the surface of the wine, or the lake, remaining flat.   If we choose the angular variation $\cos{\theta}$ for the modes the shift in the centre of mass is in the $y$ direction.  All modes with $s=1$ produce a shift in the centre of mass because they are odd in either $x$ or $y$.  For our astrophysical problem we therefore  jettison the $s=1$ modes as being unphysical.

The first non axisymmetric mode which is physical has $s=2$.  The case $j=0$ has
\begin{equation}
\zeta(r)  = a_0 \left( \frac{r}{r_e} \right )^2
\end{equation}

The next $s=2$ mode has $j=2$ and 
\begin{equation}
\zeta(r) = a_0 X^2 \left ( 1 -   \frac{3 \gamma-2}{3(\gamma-1)} X^2   \right ),
\end{equation}
with $\nu_2^2 = 4(3 \gamma-1) /(\gamma-1)$.  For this case the velocity components are
\begin{eqnarray}
v_x  & = & \frac{f \Omega^2 }{\sigma} y \left (    1 - \frac{ 3 \gamma -2}{3 (\gamma-1} \frac{3 x^2 + y^2}{r_e^2}  \right ) \\
 v_y & = &   \frac{ f \Omega^2 }{\sigma} x \left (    1 - \frac{ 3 \gamma -2}{3 (\gamma-1} \frac{  x^2 + 3y^2}{r_e^2}  \right ) 
\end{eqnarray}

\section{Exact solution for non linear, axisymmetric evolution}
\label{sec:alphaexactsolution}
In this Appendix we give the outline of the solution of the equation
\begin{equation}
\ddot{\alpha} + 6 \alpha \dot{\alpha } + 4 \alpha^3 + 4 \Omega^2 \alpha = 0.
\end{equation}
which applies to the axi-symmetric, non-linear oscillation when $\gamma=2$.

Let $V= \dot{\alpha}$ and $q = \frac12 \alpha^2$.  The equation becomes
\begin{equation}
V \frac{dV}{dq} = -(4 \Omega^2 + 6V + 8q).
\end{equation}
Define a new variable $\zeta = 4 \Omega^2 + 8q$ and substitute into the previous equation. The resulting homogeneous equation can be solved by setting $V = F \zeta$ and solving for $F$. We find
\begin{equation}
\zeta \frac{(2F+1)^2}{4F+1)} = c,
\label{eq:const}
\end{equation}
where $c$ is a constant.  Returning to the variable $V$, and after some algebra, we get
\begin{equation}
V =  \frac{d \alpha}{dt} = \frac12 \left[ c - \zeta \pm \sqrt{ c(c- \zeta)} \right].
\label{eq:dalphadt}
\end{equation}
Recalling that $\zeta = 4(\Omega^2 + \alpha^2)$  we can write $\alpha$ in terms of $\zeta$.  Next we use a new variable $\eta = \sqrt{c - \zeta}$ in place of $\zeta$ and finally set $x = \eta + \sqrt{c}$. This reduces the integral to the form 
\begin{equation}
- \int \frac{dx}{x \sqrt{ - 4 \Omega^2 + 2x \sqrt{c} - x^2} },
\end{equation}
which can be integrated and combined with the transformed (\ref{eq:dalphadt}) to give
\begin{equation}
\frac{ 8 \Omega^2 - 2x \sqrt{c} }{x \sqrt{ 4c - 16 \Omega^2} } = \sin{(2 \Omega(t + t_0)}.
\end{equation}
Transforming back to $\eta$, solving the resulting equation for $\eta^2$,  and then converting this to an equation for  $\alpha^2$ we get 
\begin{equation}
4 \alpha^2 = \frac{  (\sigma c - \sigma^2) \cos{\theta}^2 }{( \sqrt{c} + \sqrt{\sigma} \sin{\theta})^2},
\end{equation}
where $\sigma = (c - 4 \Omega^2)$, and $\theta = 2 \Omega(t + t_0)$.  Finally,
\begin{equation}
\label{ }
\alpha = \frac{ \Omega \cos{\theta} \sqrt{ c- 4 \Omega^2} }{ \sqrt{c} +   \sin{\theta}\sqrt{c- 4 \Omega^2} }.
\end{equation}
By a suitable choice of $t_0$ the cosine and sine can be interchanged.  Furthermore, from (\ref{eq:const}) it can be shown that if $\alpha$ is small then $\sqrt {c- 4 \Omega^2}$ is small. In this case the solution reduces to that given after (\ref{eq:alphalinearsol}).

\bibliography{sph}

\label{lastpage}
\end{document}